\documentclass[]{spie}  

\usepackage{amsmath,amsfonts,amssymb}
\usepackage{graphicx}
\usepackage{subcaption}
\usepackage[colorlinks=true, allcolors=blue]{hyperref}
\usepackage{gensymb}
\usepackage{amsmath}  

\title{Addressing wavelength-correlated systematics in exoplanet transmission spectroscopy: a 2D Gaussian Process approach}

\author[a]{Lokesh Manickavasaham}
\author[b]{Manjunath Bestha}
\author[b]{Sivarani Thirupathi}
\author[b]{Arun Surya}
\author[c]{Athira Unni}
\affil[a]{Indian Institute of Technology Kharagpur, Kharagpur, India}
\affil[b]{Indian Institute of Astrophysics, Bangalore, India}
\affil[c]{University of California, Santa Cruz, United States}

\authorinfo{Further author information: (Send correspondence to Lokesh Manickavasaham)\\L.M.: E-mail: lokesh.manickavasaham@gmail.com 
}

\begin{document} 
\maketitle

\begin{abstract}
Ground-based transmission spectroscopy is often dominated by systematics, obstructing us from leveraging the advantages of larger aperture sizes compared to space-based observations. These systematics could be time-correlated, uniform across all spectroscopic light curves, or wavelength-correlated, which could significantly affect the characterization of exoplanet atmospheres. Gaussian Processes were introduced in transmission spectroscopy by Gibson et al. (2012) to model correlated systematics in a non-parametric way. The technique uses auxiliary information about the observation and independently fits each spectroscopic light curve to provide robust atmospheric retrievals. However, this method assumes that the uncertainties in the transmission spectrum are uncorrelated in wavelength, which can cause discrepancies and degrade the precision of atmospheric retrievals. To address this limitation, we explore a 2D GP framework formulated by Fortune et al. (2024) to simultaneously model time- and wavelength-correlated systematics. We present its application to ground-based observations of TOI-4153b obtained using the 2-m Himalayan Chandra Telescope (HCT). As we move towards detecting smaller and cooler planets, developing new methods to address complex systematics becomes increasingly essential.  
\end{abstract}


\section{INTRODUCTION}
\label{sec:intro}  

More than 5800 exoplanets have been detected since their first discovery in 1995 (\citenum{1995Natur.378..355M}). Transmission spectroscopy is one of the most widely used methods for characterizing exoplanet atmospheres. This technique takes advantage of the transit method, where the exoplanet passes in front of its host star, obstructing a part of the starlight. By analyzing the variations in relative flux across wavelengths, we can identify the composition of an exoplanet's atmosphere. However, only a fraction of the detected exoplanets have been well characterized. The main challenge lies in the quality of the data obtained and its subsequent processing for scientific interpretation.

Ground-based transmission spectroscopy is often dominated by correlated noise, referred to as systematics, which prevents us from fully exploiting the advantages of larger aperture sizes compared to space-based observations. These systematics could be time-correlated, uniform across all spectroscopic light curves, or wavelength-correlated, which could significantly affect the results of atmospheric retrieval. Careful treatment of these systematics is essential to enhance the signal-to-noise (SNR) ratio and accurately characterize exoplanet atmospheres. Conventional methods for removing systematics involve modeling the transit light curve as a deterministic function of auxiliary observational parameters—such as airmass variations, the instrument’s position angle, the full width at half maximum (FWHM) of the spectral profile, or observing a reference star of a similar spectral type. Studies have shown that the choice of deterministic function can strongly influence the resulting transmission spectra and, consequently, the atmospheric retrievals. Gibson et al. 2012 (\citenum{2012MNRAS.419.2683G}) first introduced Gaussian Processes (GPs) to transmission spectroscopy as a more flexible approach to model the time-correlated systematics without imposing any predefined form. However, this involved modeling each spectroscopic light curve independently, assuming no wavelength correlation. This led to inconsistencies in the uncertainties of the transmission spectrum and the results of atmospheric retrievals. Fortune et al. 2024 (\citenum{2024A&A...686A..89F}) developed a two-dimensional Gaussian Process (2D GP) framework to simultaneously model both time- and wavelength-correlated systematics. In this paper, we describe the application of 2D GP to ground-based transmission spectroscopy of TOI-4153b observed using the 2-m Himalayan Chandra Telescope (HCT). Section \ref{sec:gp} provides a brief introduction to Gaussian Processes and the methodology. Section \ref{sec:data} describes the exoplanet TOI-4153b and the HCT observations along with the data reduction steps. Details of the 2D GP analysis are presented in Section \ref{sec:analysis}, followed by the results in Section \ref{sec:results} and the conclusion in Section \ref{sec:conclusion}.

\section{Gaussian Processes }
\label{sec:gp}


In probability and statistics, a Gaussian Process (GP) is a stochastic process where the random variables have a joint multivariate Gaussian (or normal) distribution. It is a supervised machine learning technique widely used for regression, classification, and function approximation tasks. It has a wide range of applications across many domains such as Machine Learning (\citenum{2006gpml.book.....R}), Astrophysics (\citenum{2012MNRAS.419.2683G}), Healthcare  and Robotics.  Mathematically, a Gaussian distribution is characterized by two parameters: the mean value ($\mu$) and the standard deviation ($\sigma$). Similarly, a GP is defined using two functions: the mean function $m(x)$ and the covariance (or kernel) function $k(x,\ x')$. A function $f(x)$ is said to be a random variable that follows a GP, if its distribution is given as:

\begin{equation}
f(x)\ \sim\ \mathcal{GP}(m(x),\ k(x,\ x'))
\end{equation}


GPs were first introduced to exoplanet transmission spectroscopy by Gibson et al. 2012 (\citenum{2012MNRAS.419.2683G}) to analyze the HST/NICMOS observations of HD-189733b. They observed that the transmission spectrum obtained using the GP method had larger uncertainties compared to those obtained using linear basis functions, as the former was able to capture the instrumental systematics more accurately. To model a light curve using GP, the mean function is the transit function ($\mathbf{T(\phi,t)}$), which depends on the transit parameters $\phi$ and time $\boldsymbol{t}$. The transit function describes the change in the flux observed as a planet transits in front of its host star. The kernel function ($\boldsymbol{\Sigma}(\boldsymbol{\theta})$) models the time-correlated systematics present in the data, which can be a function of time $\boldsymbol{t}$ and other auxiliary information collected during the observation, such as airmass variations, change in position angle and full width at half maximum (FWHM) of the spectral profile. The squared exponential kernel is widely used in transmission spectroscopy because it models smooth, slow-varying systematics over the transit time-scale, without making the model too complex (\citenum{2012MNRAS.419.2683G}). Other popular kernels include the Matérn class (3/2 and 5/2) kernels. In this work, we use the squared exponential kernel (with time $\boldsymbol{t}$ as the input parameter $x$) defined using the kernel amplitude ($h$) and the kernel length scale ($l$), and is given by:

\begin{equation}
\boldsymbol{\Sigma}(\boldsymbol{\theta})\ =\ k(x,\ x')\ =\ h^2\ \exp\left(-\frac{|x - x'|^2}{2 l^2}\right) 
\end{equation}

After initializing the GP model, the joint posterior probability ($\mathcal{P}$), which represents the probability distribution of all hyperparameters - including the transit parameters ($\boldsymbol{\phi}$) and the kernel parameters ($\boldsymbol{\theta}$) - given the observed data, is given as:

\begin{equation}
\log \mathcal{P}(\boldsymbol{\phi},\boldsymbol{\theta}|\mathbf{f}, \mathbf{X}) = \log \pi(\boldsymbol{\phi}, \boldsymbol{\theta}) + \log \mathcal{L}(\mathbf{r}|\mathbf{X}, \boldsymbol{\theta},\phi),
\label{eq:posterior}
\end{equation}

\begin{equation}
\log \mathcal{L}(\mathbf{r}|\mathbf{X}, \boldsymbol{\theta},\phi) = -\frac{1}{2} \mathbf{r}^T \boldsymbol{\Sigma}^{-1} \mathbf{r} - \frac{1}{2} \log |\boldsymbol{\Sigma}| - \frac{N}{2} \log (2\pi)
\end{equation}

where:

- $\mathcal{L}(\mathbf{r}|\mathbf{X}, \boldsymbol{\theta},\phi)$ is the likelihood that the observed data ($\mathbf{f}$) is modeled as the multivariate Gaussian distribution, 

-  \( \mathbf{r} = \mathbf{f} - \mathbf{T(\phi,t)} \) are the residuals between the observed data \( \mathbf{f}\) and the transit model \(\mathbf{T(\phi,t)}\),

- \( \pi(\boldsymbol{\phi},\boldsymbol{\theta}) \) represent the prior probability distribution over the GP hyperparameters $ (\boldsymbol{\phi},\boldsymbol{\theta})$ and

- \( \boldsymbol{X}\) represent the input parameters (here, time $\boldsymbol{t}$)



The log posterior function (Eq. \ref{eq:posterior}) is first optimized to obtain the maximum a posteriori (MAP) estimate of the parameters, which serves as an initial guess. This is followed by Markov Chain Monte Carlo (MCMC) sampling to estimate the marginal posterior distributions of each parameter.

Traditional GP analyses involve fitting each spectroscopic light curve independently assuming no wavelength correlation, but this approach has led to several discrepancies in the atmospheric retrievals. Ih \& Kempton 2021 (\citenum{2021AJ....162..237I}) identified that modeling spectroscopic light curves independently could cause correlations in the transmission spectrum and impact atmospheric retrievals. Later, Holmberg \& Madhusudhan 2023 (\citenum{2023MNRAS.524..377H}) reported the presence of wavelength-correlated systematics in JWST observations of WASP-39b and WASP-96b and tried to derive an approximate covariance matrix of the transmission spectrum. Fortune et al. 2024 (\citenum{2024A&A...686A..89F}) introduced a novel 2D GP framework to simultaneously model both time- and wavelength-correlated systematics, where the spectroscopic light curves are combined to form a 2D grid and the posterior probability function is optimized with hyperparameters shared across wavelengths. In this method, the spectroscopic light curve models are combined to form a 2D transit mean function $\boldsymbol{T}(\boldsymbol{\phi, \lambda, t})$, which depend on time $\boldsymbol{t}$ and wavelength $\boldsymbol{\lambda}$. The corresponding 2D covariance matrix, $\boldsymbol{\Sigma}_{mn}$, describes the correlation between any two data points $x_m(t, \lambda)$ and $x_n(t, \lambda)$ and is constructed using squared exponential kernels in both dimensions. A white noise component, to model uncorrelated noise, is included to the diagonal of the covariance matrix via a Gaussian term $\sigma$. The final expression is as follows:

\begin{equation}
\boldsymbol{\Sigma_{mn}}\ =\ h^2\ \exp\left(-\frac{|x_{t, m} - x_{t, n}|^2}{2 l_t^2}\right) \exp\left(- \frac{|x_{\lambda, m} - x_{\lambda, n}|^2}{2 l_{\lambda}^2}\right)\ +\ \sigma^2\ \delta_{mn}
\end{equation}

We urge the reader to refer to Fortune et al. 2024 (\citenum{2024A&A...686A..89F}) for a detailed discussion on the methodology and working of 2D GP.

\section{Data Overview}
\label{sec:data}

\subsection{TOI-4153b: A gas giant exoplanet}
\label{sec:toi}

TOI-4153b is a hot Jupiter orbiting an F-type star $(T_{eff}=6860 K)$ that was discovered in 2024 (\citenum{2024AJ....168...32S}) using the Transiting Exoplanet Survey Satellite (TESS) telescope (\citenum{2014SPIE.9143E..20R}). Located just 0.06 AU away from its host star, the exoplanet has a mass of $1.15\ M_J$ and a radius of $1.438\ R_J$. It has an orbital period of 4.6 days and an inclination of $88.75 \degree$. Stellar and planetary parameters obtained from Schulte et al. 2024 (\citenum{2024AJ....168...32S}) are summarized below. 

\begin{table}[ht]
\caption{Stellar and planetary parameters of TOI-4153b} 
\label{tab:parameter}
\begin{center}       
\begin{tabular}{cccc}
\hline
\rule[-1ex]{0pt}{3.5ex}  \textbf{Stellar Parameters} &  & \textbf{Planetary Parameters} & \\
\hline
\rule[-1ex]{0pt}{3.5ex}  Mass [$M_*/M_\odot$] & $1.572^{+0.064}_{-0.071}$ & Mass [$M_p/M_J$] & $1.15\pm0.18$ \\
\rule[-1ex]{0pt}{3.5ex}  Radius [$R_\ast/R_\odot$] & $1.605^{+0.048}_{-0.046}$ & Radius [$R_p/R_J$] & $1.438^{+0.045}_{-0.042}$ \\
\rule[-1ex]{0pt}{3.5ex}  Luminosity [$L_*/L_\odot$] & $0.713^{+0.034}_{-0.049}$ & Semimajor axis [$a$] (AU) & $0.06311^{+0.00084}_{-0.00096}$ \\
\rule[-1ex]{0pt}{3.5ex} Density [$\rho_*$] (cgs) &$0.537^{+0.044}_{-0.046}$ & Density [$\rho_p$] (cgs) & $0.477^{+0.087}_{-0.085}$\\
\rule[-1ex]{0pt}{3.5ex}  Surface gravity [$log\ g$] (cgs) & $4.224^{+0.024}_{-0.029}$ & Inclination [$i$] (deg) & $88.75^{+0.76}_{-0.6}$ \\
\rule[-1ex]{0pt}{3.5ex}  Effective Temperature [$T_{eff}$] (K) & $6860^{+150}_{-180}$ & Equilibrium Temperature [$T_{eq}$] (K) & $1669^{+29}_{-39}$ \\ 
\rule[-1ex]{0pt}{3.5ex}  Metallicity [$Fe/H$] (dex) & $0.317^{+0.100}_{-0.130}$ & Impact parameter [$b$] & $0.184^{+0.091}_{-0.110}$ \\
\rule[-1ex]{0pt}{3.5ex}  Age (Gyr) & $0.4^{+0.53}_{-0.28}$  & RV semiamplitude [$K$] ($m\ s^{-1}$) & $103\pm16$ \\
\rule[-1ex]{0pt}{3.5ex}  Distance (pc) & $419.1\pm3.5$  & Mid-transit separation [$d/R_\ast$] & $8.48\pm0.46$ \\
\hline
\end{tabular}
\end{center}
\end{table}

\subsection{HCT Observations}
\label{sec:hct}

Single-object low-resolution transmission spectroscopic observations of TOI-4153b were taken on the night of November 27, 2024 (Proposal ID: HCT-2024-C3-P25), using the Hanle Faint Object Spectrograph Camera (HFOSC) instrument mounted on the 2-m Himalayan Chandra Telescope (HCT) of the Indian Astronomical Observatory (IAO), located in the Ladakh region of the Himalayas. Low-resolution transmission spectroscopy using HCT was first demonstrated by Unni et al. 2024 (\citenum{2024MNRAS.535.1123U}). The total transit duration was 4.5 hours and 47 science frames were collected from 13:16 UT to 17:47 UT with an exposure time of 180 seconds per frame. Grism 7 was used for the observation, which has a wavelength coverage of 380 to 800 nm and a resolution of 1000. A 1340-micron width slit was used to accommodate the target. Airmass variations ranged from 1.5 (at the start of the observation) to 1.9 (at the end of the observation). 

\subsection{Data Reduction}
\label{sec:reduction}

\begin{figure}[ht]
\begin{center}
\begin{tabular}{c} 
\includegraphics[width=0.9\textwidth, height=9cm]{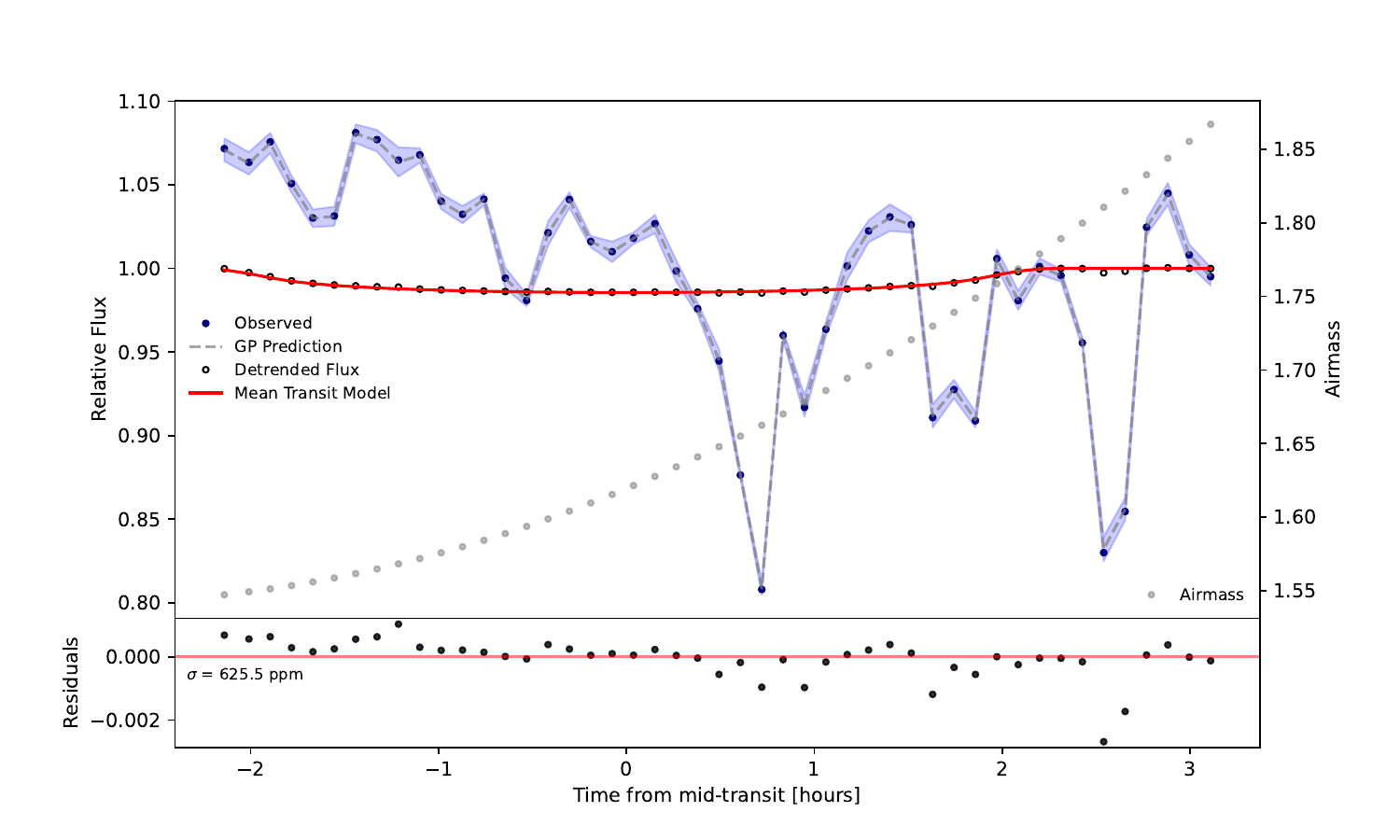}
\end{tabular}
\end{center}
\caption{White light curve analysis of TOI-4153b using 1D GP} 
\label{fig:wlc}
\end{figure}

Data reduction was carried out using a custom Python code, which involved bias subtraction, aperture extraction, and cosmic ray removal. Wavelength calibration was performed using an Fe-Ar arc lamp. The raw data was also reduced using the \texttt{IRAF} (Image Reduction and Analysis Facility) software to verify the consistency of the code, and the output spectra were similar. Spectra of wavelength range from 450 to 650 nm were integrated to construct the white light curve, avoiding the telluric region. To extract the transit parameters, the white light curve was fit with a one-dimensional GP code using \texttt{george} (\citenum{2015ITPAM..38..252A}), a Python-based package for performing Gaussian Process Regression (refer Figure \ref{fig:wlc}). For the spectroscopic light curves, the wavelength range from 400 to 700 nm was binned into 10 equal bins each of 30 nm. We then combined the spectroscopic light curves to create a 2D grid, which was the input for the 2D GP model.

\section{Data Analysis}
\label{sec:analysis}

To model spectroscopic light curves using 2D GP, we require a 2D transit mean function $\boldsymbol{T}$ and a covariance matrix $\boldsymbol{\Sigma}$. For the transit function, we adopted a quadratic limb darkening model with small planet approximation, as prescribed in Mandel \& Agol 2002 (\citenum{2002ApJ...580L.171M}). Using the Python library \texttt{JAX} (\citenum{jax2018github}), we created a vectorizing map to extend the one-dimensional transit function to two dimensions, keeping all parameters fixed excluding the planet-to-star radius ratio ($r_p/r_\ast$), the quadratic limb darkening coefficients ($u_1, u_2$), out-of-transit baseline flux ($F_{oot}$) and the gradient in baseline flux ($T_{grad}$). For the covariance matrix, we used the \texttt{LuasKernel}, which is a combination of two squared exponential functions, each for the correlation in time and wavelength. The kernel constitutes a single amplitude $h$ of correlated noise and length scales for time $l_t$ and wavelength $l_\lambda$. We included a white noise term $\sigma$ in the diagonal of the covariance matrix to account for the presence of uncorrelated noise.

The priors for the combined (mean and covariance) GP hyperparameters were selected as follows: we used normal priors for $r_p/r_\ast$ with the mean value obtained from the white light curve analysis, and normal priors for $u_1$ and $u_2$ with mean values derived from Kipping 2013 (\citenum{2013MNRAS.435.2152K}). We used uniform priors for the kernel parameters ($h$, $l_t$, $l_\lambda$ and $\sigma$). Other parameters were fixed to the values mentioned in Schulte et al. 2024 (\citenum{2024AJ....168...32S}). A summary of all the parameters (fixed and free) and the priors used in this work is given in Table \ref{tab:priors}.

\begin{table}[ht]
\caption{Summary of GP hyperparameters and priors} 
\label{tab:priors}
\begin{center}       
\begin{tabular}{ccc}
\hline
\rule[-1ex]{0pt}{3.5ex}  \textbf{Transit Parameters} & \textbf{Prior Type} & \textbf{Prior/Fixed Value}  \\
\hline
\rule[-1ex]{0pt}{3.5ex}  $r_p/r_\ast$ & Normal & $\mathcal{N}\ (0.092, 0.005)$  \\
\rule[-1ex]{0pt}{3.5ex}  $u_1$ & Normal & $\mathcal{N}\ (0.38, 0.05)$  \\
\rule[-1ex]{0pt}{3.5ex}  $u_2$ & Normal & $\mathcal{N}\ (0.20, 0.05)$  \\
\rule[-1ex]{0pt}{3.5ex}  $F_{oot}$ & Normal & $\mathcal{N}\ (1, 0.1)$  \\
\rule[-1ex]{0pt}{3.5ex}  $T_{grad}$ & Normal & $\mathcal{N}\ (0, 0.01)$   \\
\rule[-1ex]{0pt}{3.5ex}  $e$ & Fixed & $0.039$  \\
\rule[-1ex]{0pt}{3.5ex}  $a/r_\ast$ & Fixed & $8.46$  \\
\rule[-1ex]{0pt}{3.5ex}  $P$\ [days] & Fixed & $4.6174141$  \\
\rule[-1ex]{0pt}{3.5ex}  $T_{mid}$ [BJD\_TDB] & Fixed & $2460642.15$  \\
\rule[-1ex]{0pt}{3.5ex}  $i$ [deg] & Fixed & $88.75$  \\
\rule[-1ex]{0pt}{3.5ex}  $w_*$ [deg] & Fixed & $-165$  \\
\hline 
\hline 
\rule[-1ex]{0pt}{3.5ex}  \textbf{Kernel Parameters} & \textbf{Prior Type} & \textbf{Prior Value}  \\
\hline
\rule[-1ex]{0pt}{3.5ex}  $log(h)$ & Uniform & $\mathcal{U}\ (-10, 10)$  \\
\rule[-1ex]{0pt}{3.5ex}  $log(l_t)$ & Uniform & $\mathcal{U}\ (-10, 10)$  \\
\rule[-1ex]{0pt}{3.5ex}  $log(l_\lambda)$ & Uniform & $\mathcal{U}\ (-10, 10)$  \\
\rule[-1ex]{0pt}{3.5ex}  $log(\sigma)$ & Uniform & $\mathcal{U}\ (-15, -5)$  \\
\hline
\end{tabular}
\end{center}
\end{table}

We initialized the 2D GP model using \texttt{luas} (\citenum{2024A&A...686A..89F}), a Python package developed by Fortune et al. (2024). To optimize the log posterior values and find the best-fit hyperparameters, we used the Python package \texttt{PyMC} (\citenum{AbrilPla2023}), which is well suited for implementing powerful sampling algorithms such as the No U-Turn Sampler (NUTS) and advanced Markov Chain Monte Carlo (MCMC) methods like Hamiltonian Monte Carlo (HMC). We first computed the maximum a posteriori (MAP) estimate of the parameters using optimization to obtain a good initial guess for the sampler. This was followed by running 4 independent MCMC chains, each consisting of 1000 posterior draws after a tuning phase of 1000 samples.


\begin{figure}[!ht]
\begin{center}
\begin{tabular}{c} 
\includegraphics[width=0.9\textwidth, height=16cm]{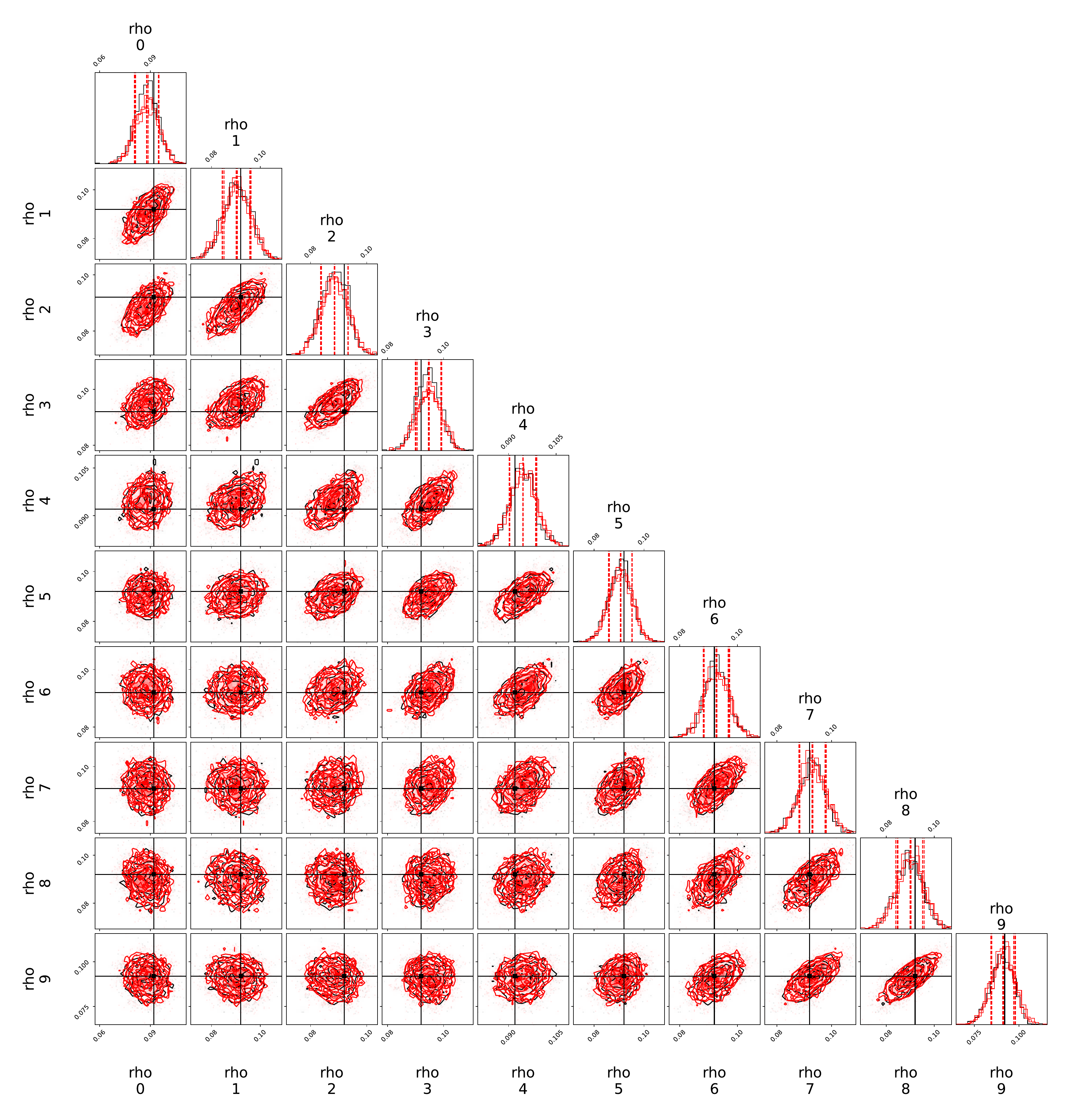}
\end{tabular}
\end{center}
\caption{Posterior distributions for $r_p/r_\ast$ (rho) from 4 independent MCMC chains. The consistent overlapping contours and quantiles across chains indicate convergence of the sampling process.}
\label{fig:corner}
\end{figure}

\section{Results and Discussion}
\label{sec:results}

\begin{figure}[!ht]
\begin{center}
\begin{tabular}{c} 
\includegraphics[width=0.98\textwidth, height=6cm]{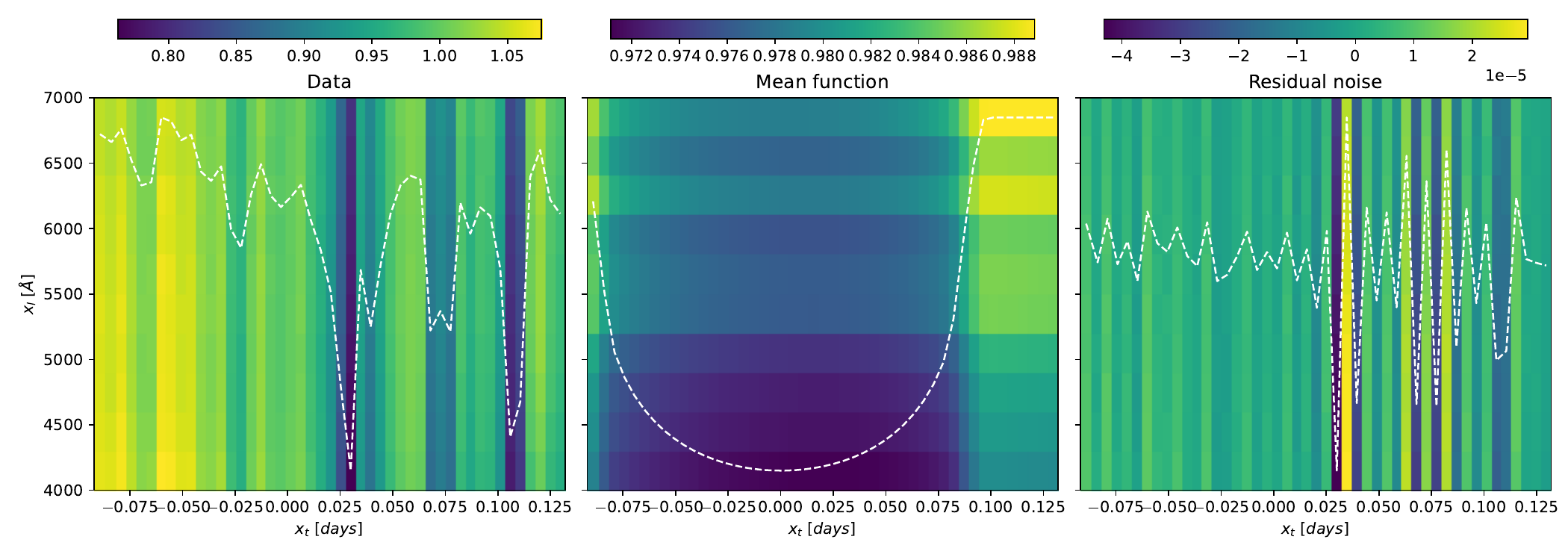}
\end{tabular}
\end{center}
\caption{2D input data with predicted 2D transit model and residual noises. In each plot, the actual data is scaled and over-plotted in white for the first spectroscopic light curve.} 
\label{fig:transit}
\end{figure}

\begin{figure}[!ht]
\begin{center}
\begin{tabular}{c} 
\includegraphics[width=0.6\textwidth, height=12.78cm]{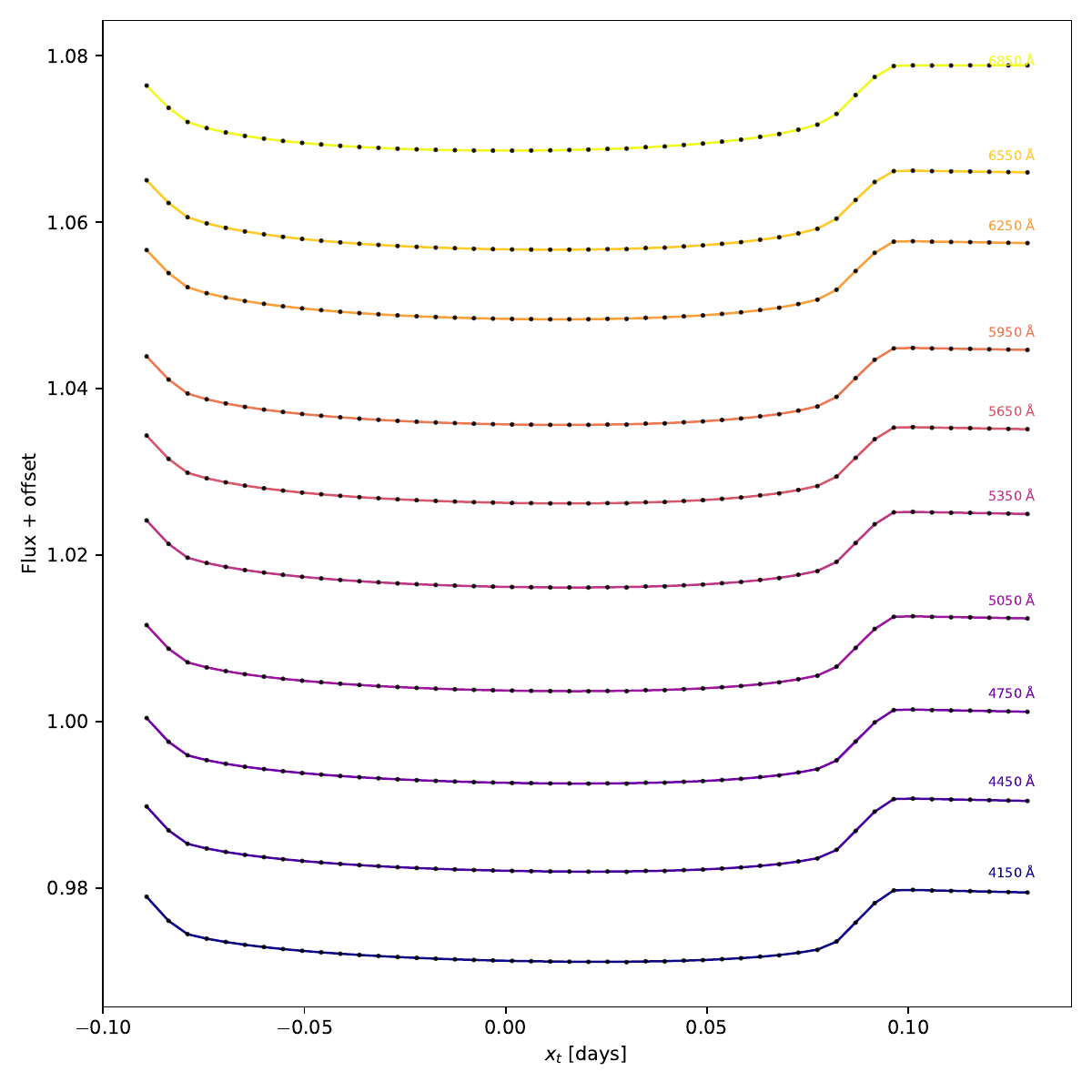}
\end{tabular}
\end{center}
\caption{Spectroscopic light curve analysis of TOI-4153b using 2D GP. The central wavelength of each spectroscopic bin is annotated above its corresponding light curve.} 
\label{fig:slc}
\end{figure}

\begin{figure}[!ht]
\centering
\begin{subfigure}[b]{0.49\textwidth}
\centering
\includegraphics[width=\textwidth, height=6cm]{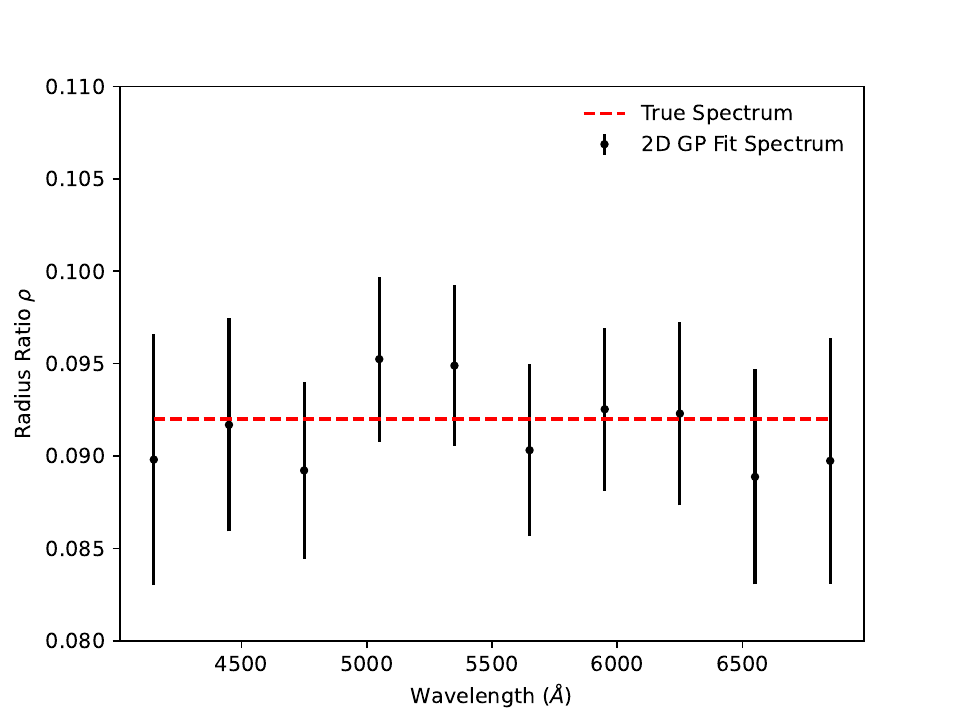}
\label{fig:spectrum1}
\end{subfigure}
\hfill
\begin{subfigure}[b]{0.49\textwidth}
\centering
\includegraphics[width=\textwidth, height=6cm]{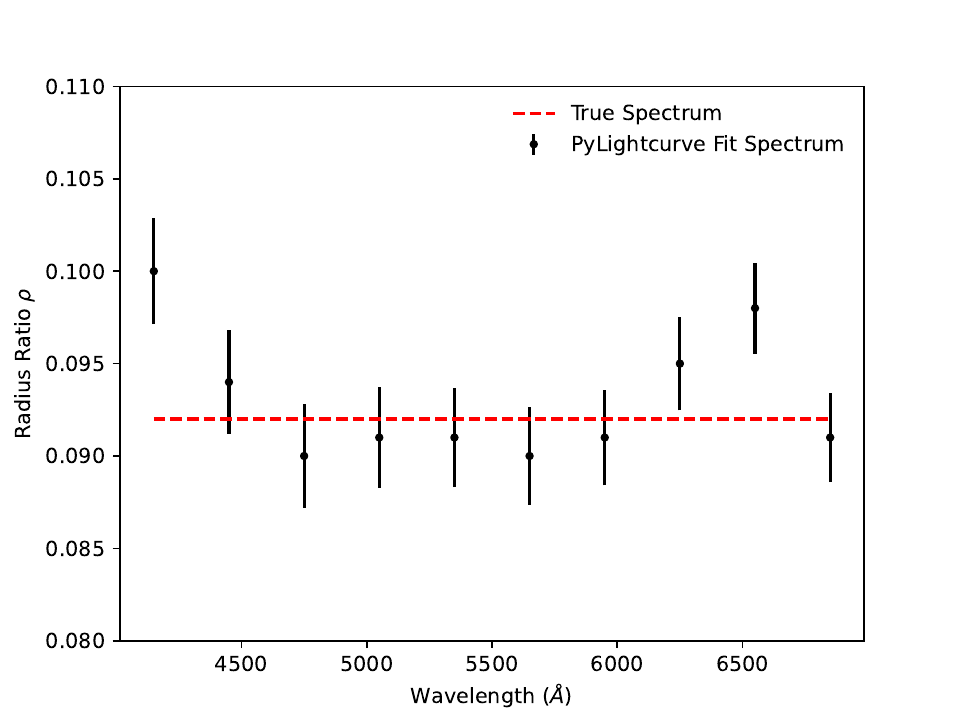}
\label{fig:spectrum2}
\end{subfigure}
\caption{Comparison between transmission spectra of TOI-4153b obtained by 2D GP analysis using \texttt{luas} (left) and by independently modeling each spectroscopic light curve using \texttt{PyLightcurve} (right)}
\label{fig:spectrum_combined}
\end{figure}

The \texttt{corner} (\citenum{corner}) plots representing the posterior distributions of the planet-to-star radius ratios ($r_p/r_\ast$) of the spectroscopic light curves and their pairwise correlations are shown in Figure \ref{fig:corner}. The plots indicate good mixing and convergence, with all chains overlapping and exploring the same regions of parameter space without divergences. The posterior distributions appear smooth, unimodal and well-sampled across most parameters.  With the optimized hyperparameters, we predicted the GP mean fit for the data and the expected 2D transit (mean) function (refer Figure \ref{fig:transit}). Using these, we removed the systematics and detrended the spectroscopic light curves, which are shown in Figure \ref{fig:slc}. Figure \ref{fig:spectrum_combined} shows the comparison between the transmission spectrum obtained using the 2D GP model and that obtained by individually fitting each spectroscopic light curve using \texttt{PyLightcurve} (\citenum{2016ApJ...832..202T}). The Transmission Spectroscopic Metric (TSM) value of an exoplanet is proportional to the expected transmission spectroscopy signal-to-noise ratio (SNR) and is used to prioritize exoplanets for atmospheric characterization (\citenum{2018PASP..130k4401K}). It estimates the signal strength in a transmission spectrum based on the planet's properties, and a higher TSM value (typically greater than $99$, \citenum{2024MNRAS.535.1123U}) indicates a better chance of detecting atmospheric features. Using the stellar and planetary parameters from Table \ref{tab:parameter}, the TSM value of TOI-4153b was found to be around 4, corresponding to a flat spectrum. 

\section{Conclusion and Future Work}
\label{sec:conclusion}

In this paper, we present the application of a two-dimensional Gaussian Process (2D GP) model to the low-resolution transmission spectrophotometric observations of TOI-4153b, obtained using the 2-m HCT. We construct a 2D grid by combining the spectroscopic light curves and use the \texttt{luas} Python package to simultaneously model both time- and wavelength-correlated systematics. We use \texttt{PyMC} to run 4 MCMC chains, each consisting of 2000 samples drawn using the NUTS algorithm. Using the optimized hyperparameters, we remove correlated systematics and detrend the light curves. We also compare the resulting transmission spectrum with that obtained by independently modeling each spectroscopic light curve using \texttt{PyLightcurve}. Given the low Transmission Spectroscopic Metric (TSM) value of TOI-4153b, a flat spectrum is expected, which is more evident in the 2D GP-fit transmission spectrum. This work can be further extended by incorporating auxiliary input parameters, such as airmass, position angle, and full width at half maximum (FWHM) of the spectral profile. This would enable more rigorous modeling of correlated systematics and lead to better atmospheric retrievals.

\appendix    


\acknowledgments 

We thank the IIA Time Allocation Committee (IIA-TAC) for providing the requested observation nights, and the staff members at the Indian Astronomical Observatory (IAO), Hanle, and the Centre for Research \& Education in Science \& Technology (CREST), Hosakote, for their assistance during the observation. In addition, L.M. gratefully acknowledges the Indian Institute of Technology Kharagpur and the Indian Institute of Astrophysics for their support and for providing the opportunity to carry out this project.

{\fontsize{11}{11}\selectfont\bibliography{report} 
\bibliographystyle{spiebib}} 
\end{document}